# Orbital-flop Induced Magnetoresistance Anisotropy in Rare Earth Monopnictide CeSb

Jing Xu[1,2], Fengcheng Wu[1,3], Jin-Ke Bao[1], Fei Han[4], Zhi-Li Xiao[1,2], Ivar Martin[1], Yang-Yang Lyu[1,5], Yong-Lei Wang[1,5], Duck Young Chung[1], Mingda Li[4], Wei Zhang[6], John E. Pearson[1], Jidong S. Jiang[1], Mercouri G. Kanatzidis[1,7], and Wai-Kwong Kwok[1]

[1]Materials Science Division, Argonne National Laboratory, Argonne, Illinois 60439, USA

[2]Department of Physics, Northern Illinois University, DeKalb, Illinois 60115, USA

[3]Condensed Matter Theory Center and Joint Quantum Institute, Department of Physics, University of Maryland, College Park, Maryland 20742, USA

[4]Department of Nuclear Science and Engineering, Massachusetts Institute of Technology, Cambridge, Massachusetts 02139, USA

[5]Research Institute of Superconductor Electronics, School of Electronic Science and Engineering, Nanjing University, Nanjing 210093, China

[6]Department of Physics, Oakland University, Rochester, Michigan 48309, USA

[7]Department of Chemistry, Northwestern University, Evanston, Illinois 60208, USA

**The charge and spin of the electrons in solids have been extensively exploited in electronic devices and in the development of spintronics. Another attribute of electrons – their orbital nature – is attracting growing interest for understanding exotic phenomena and in creating the next-generation of quantum devices such as orbital qubits. Here, we report on orbital-flop induced magnetoresistance anisotropy in CeSb. In the low temperature high magnetic-field driven ferromagnetic state, a series of additional minima appear in the angle-dependent magnetoresistance. These minima arise from the anisotropic magnetization originating from orbital-flops and from the enhanced electron scattering from magnetic multidomains formed around the first-order orbital-flop transition. The measured magnetization anisotropy can be accounted for with a phenomenological model involving orbital-flops and a spin-valve-like structure is used to demonstrate the viable utilization of orbital-flop phenomenon. Our results showcase a contribution of orbital behavior in the emergence of intriguing phenomena.**

**Introduction**

The shape of the electron cloud in a solid is denoted as an orbital [1]. The orbital degree of freedom and its interplay with the charge and spin of the electron are of importance for understanding electronic phenomena [1-3]. In transition-metal oxides, the ordering and correlation of the anisotropic d-orbitals play a critical role in the metal-insulator transition and the occurrence of colossal magnetoresistance [1,2]. Orbital excitations were found to be the key in understanding the temperature dependence of the resistivity and optical conductivity in the ferromagnetic phase of doped manganites [4]. The multiferroicity in $GeV_4S_8$ is thought to originate from an orbital ordering that reorganizes the charge within the transition metal clusters [5]. Furthermore, electron orbitals are considered to be the origin of the extremely anisotropic superconducting gap in the nematic phase of FeSe superconductors [6]. Although less explored than spins, orbitals are attracting increasing attention in quantum information science and applications [7]. For example, orbital angular momentum has been proposed for quantum information processing [8,9]. Controls of orbital states using electrical field [10], light [11], and lattice strain [7] have also been attempted, aiming at potential applications such as orbital qubits.

The recent quest for the origin of the extremely large magnetoresistance (XMR) in rare earth monopnictide CeSb demonstrated that the Ce orbital state is partly responsible for the XMR [3]. Here, we reveal an orbital-flop transition that induces anisotropic magnetoresistances and magnetizations in CeSb. We found that the anisotropy of the magnetoresistance in CeSb in the paramagnetic phase resembles that of the orbitally-quenched compound GdBi. However, unlike GdBi, when CeSb enters the ferromagnetic phase at low temperatures, it shows additional extrema in the field-orientation dependence of the magnetoresistance. We attribute the additional anisotropy of the magnetoresistance in CeSb at high magnetic fields to an abrupt reorientation by 90° in the Ce $\Gamma_8^{(1)}$ orbitals. This hypothesis is supported by the field-orientation dependence of the critical magnetic field above which CeSb is ferromagnetic. We validate this mechanism by accounting for the anisotropy of experimentally obtained magnetizations with a phenomenological model involving such an orbital-flop. We further construct a spin-valve-like structure consisting of CeSb and permalloy (Py, $Ni_{80}Fe_{20}$) to directly elucidate the orbital-flops through an abrupt change in the field-orientation dependence of the tunneling magnetoresistance between the Py layer and the CeSb crystal. This work reveals a new type of orbital-induced intriguing property, which can stimulate further exploration of materials with anisotropic orbitals. It provides direct evidence of electron orbital contributions to the observed XMR. Our results also demonstrate the potential of utilizing orbital-flop transitions in electronic applications.

**Results**

**Orbital-flop induced magnetoresistance anisotropy**

CeSb is one of the rare-earth monopnictides LnX (e.g. Ln = Y, Sc, Nd, Lu, La, Ho, Ce and X = Sb/Bi) that have recently been investigated intensively in the context of topological materials [11-19] and XMR phenomenon [3,20-31]. While its topological nature is currently under debate [3,16-18], its magnetoresistance is the largest among the rare-earth monopnictides, reaching up to $1.6\times10^6$ % at a magnetic field of 9 T [3]. Distinct from nonmagnetic La and Y monopnictides, CeSb with the Ce 4*f*-electron state shows various long periodic magnetic structures at low temperatures even in zero magnetic field [3,32,33] (Supplementary Figure 1). In the presence of a magnetic field, it exhibits at least 14 magnetic phases [3,33] (also see Supplementary Figure 1). The tunable magnetic structures provide an opportunity to investigate the possible correlation between its purported topological property and magnetism [3,18]. Recently, Ye et. al. revealed the magneto-orbital control of XMR in CeSb [3], where they presented magnetoresistance measurements in a fixed-field orientation (along [001]). In this work we reveal properties of CeSb resulting from varying the magnetic-field orientation.

We studied three crystals from the same batch, with Sample A in a standard four-contact magneto-transport configuration and Sample B in two-contact spin-valve-like structure for resistance measurements, and contact-less Sample C for magnetization characterizations. Figure S2a shows the temperature dependent resistance curve $R(T)$ for Sample A with typical zero-field magnetic ordering at $T_N \approx 16$ K. Figure S2b shows the magnetoresistance $MR(H) = [R(B)-R_0]/R_0$ at $T = 3$ K and **H** ∥ [001], which yields a large value of ~$1.1 \times 10^4$ % at $\mu_0 H = 9$ T. Here, $R(B)$ and $R_0$ are the resistance with and without magnetic field, respectively. The $MR(H)$ also shows an apparent kink at ~3.62 T associated with an antiferromagnetic to ferromagnetic transition with increasing field, consistent with those reported in the literature [3,18], indicating the high quality of our crystals.

CeSb has a rock-salt cubic crystal lattice, resulting in identical electronic structures along the $k_x$, $k_y$ and $k_z$ directions of the Brillouin zone. Similar to LaSb, it has a bulk anisotropic Fermi surface with elongated electron Fermi pockets [17]. Thus, we expect to see anisotropic magnetoresistance in CeSb when the magnetic field is tilted away from the [001] direction. As shown in Supplementary Figure 3b for the data obtained at $T = 25$ K and at various magnetic fields, we indeed observed anisotropic magnetoresistances with a four-fold symmetry in the paramagnetic state, which as in LaSb [24], is probably associated with the anisotropic bulk Fermi surface. However, in the low-temperature magnetically ordered states, the anisotropy of the magnetoresistance exhibit additional minima as presented in Figs.1b and 1c for the results obtained at $T = 3$ K and Supplementary Figure 4 for $T = 5$ K. Namely, the $R(\theta)$ curves for $\mu_0 H \geq 5$ T exhibits additional minima along <011> directions, i.e., at $\theta = 45°$, 135°, 225° and 315°, which are more pronounced than the minima at $\theta = 0°$, 90°, 180° and 270° expected from the bulk Fermi surface.

Furthermore the $R(\theta)$ curves for $\mu_0H \leq 4$ T show additional complicated features. For example, the minima at $\theta = 90°$ and $270°$ in the curve for $\mu_0H = 4$ T become barely recognizable. The curves for $\mu_0H = 4$ T to 2 T show additional features at $90° < \theta < 135°$ and $270° < \theta < 315°$. Moreover, anomalies at <011> persists down to 2 T. These results clearly reveal the contribution of magnetic structures to the magnetoresistance and its anisotropy.

The magnetism in CeSb arises from the Ce $4f^1$ electrons with $\Gamma_7$ and $\Gamma_8^{(1)}$ crystal field eigenstates, which have magnetic moments of $0.71\mu_B$ and $1.57\ \mu_B$, respectively, with $\mu_B$ being the Bohr magneton [32]. In the high-temperature paramagnetic phase, $\Gamma_7$ is the preferred orbital state while in the low-temperature magnetically-ordered phase, $\Gamma_8^{(1)}$ becomes energetically favored, as depicted in the color map of the $\Gamma_8^{(1)}$ occupation recently reported by Ye et. al. for an external magnetic field applied in the [001] direction [3]. For the magnetic field along the [001] and [010] directions in Fig.1a, the planarly shaped $\Gamma_8^{(1)}$ orbital state induces a magnetic moment **M**$_0$ parallel to the magnetic field. When the magnetic field is tilted away from the [001] and [010] directions there are two possible scenarios for the relationship between **H** and **M**: (1) **M** rotates with **H**, yielding an angle-independent induction $B_\theta = \mu_0H+M_0$. In this case the magnetic effects may not contribute to the anisotropy of the magnetoresistance. (2) **M** suddenly rotates 90°, i.e., flops, when **H** crosses the <011> direction, as schematically presented in Fig.1d. For example, when **H** is applied at $\theta < 45°$, **M** from $\Gamma_8^{(1)}$ orbitals is along [001] with a constant value $M_0$. Once $\theta$ is larger than 45° (but < 135°) **M** becomes parallel to [010] while its magnitude remains the same. Namely, the $\Gamma_8^{(1)}$ orbitals and the induced **M** flop when the orientation of **H** crosses $\theta = 45°$. The same **M**- and orbital-flops occur at $\theta = 135°$, 225° and 315°. This leads to an angle-dependent induction $B_\theta = [(\mu_0H)^2 + (M_0)^2 + 2\mu_0HM_0 cos\varphi]^{1/2}$, where $|\varphi| \leq 45°$ is the angle between **H** and **M** and its relationship to $\theta$ is $\varphi = \theta - n\pi/2$ with $n = $ 0, 1, 2, 3, and 4 for $0° \leq \theta \leq 45°$, $45° \leq \theta \leq 135°$, $135° \leq \theta \leq 225°$, $225° \leq \theta \leq 315°$, and $315° \leq \theta \leq 360°$, respectively. Hence, $B_\theta$ has a minimum at **H** ∥ <011>, resulting in minima in magnetoresistance that depends positively on magnetic induction, as revealed by the $R(H)$ curve shown in Supplementary Figure 2b for **H** ∥ [001]. Thus, orbital-flop proposed in Fig.1d can contribute to the observed minima or anomalies in $R(\theta)$ curves at $\theta = 45°$, 135°, 225° and 315° in Fig.1b and 1c.

To better understand the angle dependence of the magnetoresistance we also need to know the anisotropy contributed by the bulk Fermi surface besides the component from orbital-flops. The results in Fig.1 and Supplementary Figure 4 show that the MRs in CeSb for **H** ∥ **b** ($\theta = 90°$ and 270°) are nominally smaller than those for **H** ∥ **c** ($\theta = 0°$ and 180°), particularly in the $R(\theta)$ curves for 1 T < $\mu_0 H$ < 4 T. This behavior is unlike LaSb [20], which show symmetric MR values for the same four **H** orientations as expected for a cubic lattice. A symmetry-breaking in CeSb can be induced by a small (of the order of $10^{-3}$) tetragonal distortion at $T < T_N$, which was observed by Hullinger et al [34]. A small mis-alignment between the magnetic field rotation plane and the current flow direction can also cause significant asymmetry in the magnetoresistance anisotropy in the XMR state [35]. Thus, it is difficult to make a quantitative analysis. For a qualitative comparison with the experimental data, we present in Supplementary Figure 5 the calculated $R(\theta)$ curve including both the contributions from the bulk Fermi surface and the orbital-flops. We chose $R(\theta)$ curve at a high magnetic field ($\mu_0 H = 7$ T) where CeSb is ferromagnetic phase with a known $M_0$ and remains in this phase when changing magnetic field orientations. Although the suppression of $R(\theta)$ due to the orbital-flops is clearly identifiable for **H** ∥ <011>, the calculated results show a much weaker effect than the experimentally observed ones, particularly at angles close to $\theta = 45°$, 135°, 225° and 315°. It is possible that multi-domain states exist in the orbital-flop region, similar to that in spin-flop materials in which multi-domains were observed in the vicinity of the first-order spin flop transition, where local flops occur due to thermal excitation and inhomogeneities such as defects in the crystals [36,37]. If so, the electron mobility μ decreases due to scattering at the domain walls, resulting in further reduction of the magnetoresistance that typically follows the relationship of $MR \sim \mu^2$ when the Hall factor $\kappa_H \ll 1$ (see Supplementary Figure 2b) [38].

In the scenario of orbital-flops, the magnetic moment **M** orients along either the **b**- or **c**-axis in Fig.1d and the magnetic phase transition is induced by the component $\mathbf{H}_M$ of the magnetic field **H** along **M**. That is, the magnetic field at which the phase transition occurs should have a $1/cos\varphi$ dependence. Figure 2a presents $R(H)$ curves at $T = 3$ K and for $\theta = 90°$, 135° and 180°. It is evident that the ferromagnetic phase transition field $H_{FM}$ for $\theta = 135°$ (**H** ∦ **M**, $\varphi = 45°$) is larger than those for $\theta = 90°$ and 180° (**H** ∥ **M**, $\varphi = 0°$). Figure 2b presents additional data of $H_{FM}$ for $90° \leq \theta \leq 270°$. It shows that the anisotropy of $H_{FM}$ indeed follows $1/cos\varphi$ very well. We note that in Fig.2a the $R(H)$ curves for theoretically equivalent **H** ∥ **b** and **H** ∥ **c** do not overlap and the values of $H_{FM}$ in these two orientations also differ from each other. Furthermore, the $R(H)$ curve for **H** ∥ **b** in Fig.2a even exhibits an additional kink at $H_{FM*}$. These results again reveal possible lattice distortion at low temperatures and small sample mis-alignment.

Figure 1 and Supplementary Figure 4 show the reduction in the magnitude of the minima at **H** ∥ <011> with decreasing magnetic field. In Fig.2d we present $R(\theta)$ curves at $\mu_0 H = 7$ T and at various temperatures. With increased temperatures, the minima at **H** ∥ <011> become shallower and eventually disappear at $T \sim 10$ K. In Fig.2a we can identify a magnetic field $H_{CR}$ above which the magnetoresistance along **H** ∥ <011> is less than that of **H** ∥ **b** or **H** ∥ **c**. Its temperature dependence, shown in Fig.2c, defines a magnetic field and temperature regime in which the minima in the $R(\theta)$ curves at **H** ∥ <011> are apparent. By comparing it with the temperature dependence of $H_{FM}$ presented in the same figure, we conclude that the effects of orbital-flops on the anisotropy of the magnetoresistance is most pronounced in the ferromagnetic phase. This can be understood if the minima/anomalies at **H** ∥ <011> are determined by the amplitude of **M** and the electron scattering by domain walls, both of which are largest in the ferromagnetic phase.

**Orbital-flop induced anisotropic magnetization**

To further confirm the bulk orbital-flop mechanism, we investigate the angle dependence of the magnetization of a CeSb crystal (Sample C) using a magnetometer that measures the magnetization component along the magnetic field direction. Following the magnetic field configuration as in Sample A, the measured magnetization is expected to follow $M = M_0\cos\varphi$, where $\varphi = \theta - n\pi/2$ as defined above and $\mathbf{M}_0$ aligns only along **b** and **c**-axes, as represented by arrows in Fig.1d. In other words, the angle dependence of the magnetization $M(\theta)$ should exhibit minima at $\varphi = 45°$, i.e., $\theta = 45°$, 135°, 225° and 315°. Figure 3a presents $M(\theta)$ curves obtained at $T = 5$ K and at various magnetic fields in the angle range of $0° \leq \theta \leq 180°$. Minima can be clearly identified at $\theta = 45°$ and 135° at $\mu_0H = 5$ T and 6 T. Similar minima occur in the $M(\theta)$ curves for $\mu_0H = 7$ T and in temperatures ranging from 5 K to 25 K as presented in Fig.3b. However, some of the curves, e.g., those at $T = 5$ K and $\mu_0H = 4$ T in Fig.3a as well as at $T = 15$ K and $\mu_0H = 7$ T in Fig.3b clearly deviate from a cosine relationship. In those curves the ratios $\gamma$ of the moments at $\theta = 0°$, 90° and 180° to those at $\theta = 45°$ and 135° are also larger than the expected value $\gamma = 1/cos45° = \sqrt{2}$. These deviations simply arise from the different magnetic states of the crystal when the magnetic field is rotated, as indicated by the magnetic field dependence of the moment at $T = 5$ K and at different field orientations as shown in Supplementary Figure 6 and the temperature dependence at $\mu_0H = 7$ T in Fig.3c. Similarly, the smaller values of $\gamma$ for $M(\theta)$ curves at $\mu_0H \leq 3$ T in Fig.3a and $T \geq 20$ K in Fig.3b are also understandable if the crystal is not entirely in the pure ferromagnetic state and the residual paramagnetic components do not contribute to the magnetic anisotropy. The phase diagram of $H_{FM}$ for the ferromagnetic transition at $\theta = 135°$ in Fig.2c defines the condition under which the CeSb crystal will in the ferromagnetic state for all magnetic field orientations. For example, ferromagnetic phase should exist up to $T \sim 10$ K at $\mu_0 H = 7$ T, independent of the magnetic field orientations. The $M(\theta)$ curves obtained in this magnetic field-temperature regime indeed follow a cosine-like relationship, e.g. those for $T = 5$ K and 10 K in Fig.3b. As presented in Fig.3c for $\mu_0 H = 7$ T, the values of $\gamma = 1.263 \pm 0.015$ are nearly constant up to $T = 10$ K in the ferromagnetic phase, albeit smaller than the expected value of $\sqrt{2}$.

To elucidate the magnetization anisotropy, we constructed a phenomenological model to describe the ferromagnetic phase of CeSb. We assume that the Ce $f$-electrons occupy the same state at all sites and consider only the nearest-neighbour ferromagnetic coupling. Strong spin-orbit coupling splits the $f$ states into total-angular-momentum $J = 5/2$ and $J = 7/2$ multiplets and the crystal field further splits the $J = 5/2$ multiplet into two groups, $\Gamma_7$ doublet and $\Gamma_8$ quartet. The $\Gamma_8$ states can be described as $\left|\Gamma_8^{(1),\pm}\right\rangle = \sqrt{\frac{5}{6}}\left|\pm\frac{5}{2}\right\rangle + \sqrt{\frac{1}{6}}\left|\mp\frac{3}{2}\right\rangle$ and $\left|\Gamma_8^{(2),\pm}\right\rangle = \left|\pm\frac{1}{2}\right\rangle$. For a magnetic field along the **c**-axis, $\left|\Gamma_8^{(1)}\right\rangle$ is the ground state. This leads to the mean-field Hamiltonian:

$$\mathcal{H}_{FM} = \sum_{\mathbf{R}}[g\mu_B \mathbf{H} - \varepsilon \langle \tilde{\mathbf{J}}_{\mathbf{R}} \rangle] \cdot \tilde{\mathbf{J}}_{\mathbf{R}} \qquad (1)$$

where **R** labels Ce sites and $\tilde{\mathbf{J}}$ is the three-dimensional angular momentum operator that is projected onto $\Gamma_8$ quartet as accentuated by the tilde (~). The Hamiltonian $\mathcal{H}_{FM}$ in Eq.1 incorporates the cubic anisotropy due to the projection. $g = 6/7$ is the effective g-factor for $J = 5/2$ multiplet. **H** is the applied magnetic field. $\langle \tilde{\mathbf{J}}_{\mathbf{R}} \rangle$ is the average value with respect to the ground state. $\varepsilon > 0$ is the effective ferromagnetic coupling constant, which favors a state with $\langle \tilde{\mathbf{J}}_{\mathbf{R}} \rangle$ along any one of the crystal axes, **a**, **b** or **c**. On the other hand, the applied magnetic field favors a state with $\langle \tilde{\mathbf{J}}_{\mathbf{R}} \rangle$ antiparallel to **H**. Therefore, there can be a first-order phase transition as **H** rotates, for example, from **c** to **b** axis. By solving $\mathcal{H}_{FM}$ using iterations, we obtain the magnetization ($\mathbf{M} = -g\mu_B \langle \tilde{\mathbf{J}}_\mathbf{R} \rangle$) as a function of the magnetic field orientation angle $\theta$. We find the derived $M(\theta)$ curve depends on the coupling constant $\varepsilon$ (see Supplementary Figure 7). At $\varepsilon/(g\mu_B B) = 2.15$, the calculated $M(\theta)$ curves closely follow the experimental data (see solid curves in Fig.3b). This gives $\gamma = 1.314$, which is close to the experimental value of 1.263. Our model considers a perfectly ordered magnetic structure. It does not include the effects of multidomains, which can be the cause for slight differences in the shape of the $M(\theta)$ curves before and after the orbital-flops (see data around $\theta = 45°$ and $135°$ for the curve obtained at $\mu_0 H = 6$ T in Fig.3a and the curves for 5 K and 10 K in Fig.3b). Figure 3d presents the angle dependence of the corresponding **M** components along **b**- and **c**-axis. Besides the expected sudden changes in the values of $\mathbf{M_b}$ and $\mathbf{M_c}$ induced by the orbital-flop at $\theta = 45°$, the results indicate that $M(\theta) = M_c\cos\varphi + M_b\sin\varphi$, with $M_c \approx M_0$, and $\varphi = \theta$ for $0° < \theta < 45°$; $M(\theta) = M_b\cos\varphi + M_c\sin\varphi$ with $M_b \approx M_0$ and $\varphi = 90° - \theta$ for $45° < \theta < 90°$. Since $M_b \neq 0$ for $0° < \theta < 45°$ and $M_c \neq 0$ for $45° < \theta < 90°$, the measured $M(\theta)$ values will be larger than those calculated from $M = M_0\cos\varphi$ (except at $\varphi = 0$) and $\gamma < \sqrt{2}$ accordingly.

**Results from orbitally quenched GdBi**

To highlight the important role of orbital flops on the observed anisotropy in the magnetoresistance and magnetization, we compare our results to those of GdBi, which is similar to CeSb in electronic and magnetic structures, but is orbitally quenched. A similar comparison was conducted by Ye et al [3] for **H** ∥ **c** to demonstrate the importance of the *f*-orbital degree of freedom on the occurrence of XMR. Here, we carry the comparison further by reporting on the angular dependence of the resistivity, $R(\theta)$, as shown in Figure S8a obtained from a GdBi crystal at $T = 5$ K and $\mu_0 H = 9$ T to 1 T in intervals of 1 T. Clearly, no minima or anomalies occur at **H** ∥ <011>, i.e., $\theta = 45°, 135°$, 225° and 315° (see Fig.1b, Fig.1c and Supplementary Figure 4 for comparison). Similarly, the $M(\theta)$ curves shown in Supplementary Figure 8b for $\mu_0 H$ = 7 T to 1 T in intervals of 2 T indicate that GdBi is magnetically isotropic. These comparisons clearly reveal the role of orbitals on the anisotropy of the magnetoresistance and the magnetization observed in CeSb.

**Orbital-flop based spin-valve-like structure**

We fabricated a spin-valve-like structure to demonstrate the functionality of orbital-flop phenomenon. A spin valve comprising of two ferromagnetic electrodes can have high- and low-resistance states, depending on the relative orientations of the magnetic moments in the electrodes. Spin valve structures have been used in microelectronic devices such as hard-drive read heads and magnetic random access memories [39]. They can also serve as a useful platform for probing new phenomena such as triplet Cooper pairs [40]. As presented in Fig.4a we devised a spin-valve-like structure by depositing a 30 nm permalloy (Py) layer (covered with 20 nm thick Au) onto the surface of a CeSb crystal and attaching gold wires using silver paste. In this two-contact structure, the measured resistance includes both the component $R_{CeSb}$ of the CeSb crystal and the tunneling resistance $R_{SV}$ between the Py layer and the CeSb crystal at the two contacts, as described by the equivalent circuit in Fig.4b. We followed the same measurement procedure used in Sample A, i.e., rotating the magnetic field in the (100) plane (see the schematic in Fig.4a for defining the angle $\theta$). The orbital-flops should produce a two-value tunneling resistance state, $R_{sv}$, with a low- and a high-resistance state corresponding to the parallel and perpendicular configurations of the magnetic moments in the Py layer and in the CeSb crystal, respectively (see Fig.4c for schematics). These two resistance states switch at $\theta$ = 45°, 135°, 225° and 315° at which orbital-flops occur. Since $R_{SV}$ is in series with $R_{CeSb}$, the measured total magnetoresistance of this spin-valve-like structure should exhibit a stronger asymmetry in the angle dependence than the conventional four-contact configuration of Sample A. In fact, this is exactly what we observed: Comparing to Fig.1 and Supplementary Figure 4 for Sample A, the angle dependence of the magnetoresistance in Fig.4d for Sample B with the spin-valve-like structure has higher asymmetry in the ferromagnetic state, with significant suppression of the magnetoresistances at $45° < \theta < 135°$ and $225° < \theta < 135°$ relative to those at $0° \leq \theta < 45°$, $135° < \theta < 225°$ and $315° < \theta \leq 360°$.

**Discussion**

Our results reveal rich physics in CeSb, such as magnetic and orbital modulations of XMR, in addition to the orbital-driven anisotropy in the magnetoresistance and magnetization. Our approaches can be readily applied to explore orbital induced phenomena in other cerium monopnictides, which also have numerous magnetic phases and magnetoresistance behaviours [41,42]. Recent x-ray angle-resolved photoemission spectroscopy (ARPES) investigations show that cerium monopnictides can change from topologically trivial to nontrivial, with the increase of spin-orbit coupling as in CeP and CeBi respectively [43]. Thus, comparative studies on different cerium monopnictides using our approach can shed light upon the role of orbitals in achieving topological states and in uncovering topological properties originating from orbitals.

Besides CeSb, recent pursuit of the XMR phenomenon has led to the exploration on other rare-earth monopnictides including GdBi [3], NdSb [23,26], DySb [27], and HoBi [30,31]. These materials not only exhibit XMRs but also are magnetic, with localized $f$ electron moments. With the exception of Gd monopnictides in which Gd $4f$ orbitals are exactly half occupied and the orbital angular momentum is zero, $f$ orbital moments generally cannot be quenched by the crystal field [44]. Therefore, orbital-driven phenomena including anisotropic magnetoresistances and magnetizations are expected to be observed in these magnetic rare-earth monopnictides. Since orbital component dominates the total magnetic moment, orbitals may play an important role in the XMR of HoBi, which only occurs in the ferromagnetic phase at high enough magnetic fields [31].

This work unambiguously demonstrates the existence of orbital-flops, i.e., rotation of 90° in orbital orientation, in CeSb. Our phenomenological model also shows that spin-flops accompany orbital-flops due to strong spin-orbit coupling. The spin-flops in CeSb is induced by rotating magnetic field and differs from the classic spin-flops widely observed in antiferromagnets [36,37,45-48]. The latter is a spin re-orientation transition when the strength of the magnetic field changes while its direction is fixed. When the spin-charge coupling is strong, classic spin-flops can also induce anisotropic magnetoresistance [45].

Our investigations on anisotropic properties can also shed light on some outstanding issues. For example, the observed phase transition at $H_{FM^*}$ for **H** ∥ **b** while absent for **H** ∥ **c** in our CeSb crystal provides an explanation of the inconsistency in the magnetic phase transitions in the magnetic field range from 3.5 T to 4.0 T, with two transitions reported by Ye et al [3] and only one observed by Wiener et al [33].

**Methods**

**Sample preparation.** CeSb single crystals were grown with Sn flux as described in Ref.49. Ce powder (99.9%, Alfa Aesar), Sb pieces (99.999%, Plasmaterials Inc) and Sn shots (99.999%, Alfa Aesar) with an atomic ratio of Ce:Sb:Sn = 1:1:20 were loaded into an alumina crucible. The alumina crucible, covered with a stainless steel frit, was sealed in a 13mm diameter quartz tube under a vacuum of $10^{-3}$ mbar. The quartz tube was heated up to 1150 °C in 20 h and kept at that temperature for another 12 h, then cooled down to 800 °C at a rate of 2 °C/h. Finally, the tube was centrifuged at 800 °C to separate CeSb crystals from the Sn flux. Large crystals with dimensions up to 5×5×4 mm³ were harvested. The crystals were characterized by single crystal X-ray diffraction on the diffractometer STOE IPDS 2T and a rock-salt type structure of CeSb was confirmed for those crystals. GdBi single crystals were grown out of Bi flux as shown in Ref.49. Gd pieces (99.9%, Alfa Aesar) and Bi shots (99.999%, Sigma-Aldrich) with an atomic ratio of Gd:Bi = 1:4 were loaded directly into an alumina crucible. The alumina crucible was sealed in a 15 mm quartz tube under a vacuum of $10^{-3}$ mbar. A stainless steel frit was covered and fixed on the top of the alumina crucible. The tube was heated up to 1150 °C for 20 h, held at that temperature for 12 h, and then cooled down to 940 °C at a rate of 2 °C/h, where the tube was centrifuged to separate the crystals from the Bi flux. Crystals with a typical dimension of $2 \times 2 \times 3$ mm$^3$ were harvested. The crystals obtained were confirmed to be the GdBi phase by single crystal X-ray diffraction.

**Resistance measurements.** We conducted DC resistance measurements on CeSb crystals in a Quantum Design PPMS-9 using constant current mode ($I = 1$ mA). The electric contacts were made by attaching 50 µm diameter gold wires using silver epoxy, followed with baking at 120 °C for 20 minutes. Angle dependence of the resistance was obtained by placing the sample on a precision, stepper-controlled rotator with an angular resolution of 0.05°. The inset of Figure 1 shows the measurement geometry where the magnetic field **H**($\theta$) is rotated in the (100) plane and the current **I** flows along the [100] direction, such that the magnetic field is always perpendicular to the applied current **I**. We define the magnetoresistance as $MR = [R - R_0]/R_0$, where $R$ and $R_0$ are resistivities at a fixed temperature with and without magnetic field, respectively.

**Magnetization measurements.** We measured the angular and temperature dependences of the magnetization of a CeSb crystal using a Quantum Design MPMS XL equipped with a horizontal sample rotator with an angular resolution of 0.1° and reproducibility of ±1°.

**Data availability.** The data that support the findings of this study are available from the corresponding authors upon reasonable request.

ACKNOWLEDGEMENTS

Crystal growth, magnetotransport measurements and theoretical investigation were supported by the U.S. Department of Energy, Office of Science, Basic Energy Sciences, Materials Sciences and Engineering. W. Z. acknowledges support from U.S. National Science Foundation under Grant No. DMR-1808892 and the DOE Visiting Faculty Program. Work of F. W. at University of Maryland is supported by Laboratory for Physical Sciences. Y. L. W and Y. Y. L acknowledge supports by the National Natural Science Foundation of China (61771235 and 61727805) and the National Key R&D Program of China (2018YFA0209002).

**Author contributions**

Z. -L. X. and W. Z. designed the experiments; J. -K. B., D. Y. C. and M. G. K. grew the crystals; J. X., J. E. P., J. S. J. and Y. -L. W. conducted magnetoresistance and magnetization measurements; J. X., F. H., M. L., and Y. -Y. L. contributed to data analysis; F. W. and I. M. conducted theoretical investigations; Z. L. X., W. Z., and W. -K. K. wrote the paper. All of the authors reviewed the manuscript.

**Additional information**

**Competing interests:** The authors declare no competing interests.

**Materials & Correspondence:** xiao@anl.gov; martin@anl.gov; weizhang@oakland.edu

# References

1. Tokura Y. & Nagaosa, N. Orbital physics in transition-metal oxides. *Science* **288**, 462-468 (2000).

2. Streltsov, S. V. & Khomskii, D. I. Orbital physics in transition metal compounds: new trends. *Physics - Uspekhi* **60**, 1121-1146 (2017).

3. Ye, L., Suzuki, T., Wicker, C. R. &. Checkelsky, J. G. Extreme magnetoresistance in magnetic rare-earth monopnictides. *Phys. Rev. B* **97**, 081108(R) (2018).

4. van den Brink, J., Horsch, P., Mack, F. & Oles, A. M. Orbital dynamics in ferromagnetic transition-metal oxides. *Phys. Rev. B* **59**, 6795-6805 (1999).

5. Singh, K. et al. Orbital-ordering-driven multiferroicity and magnetoelectric coupling in $GeV_4S_8$. *Phys. Rev. Lett.* **113**, 137602 (2014).

6. Liu, D. et al. Orbital origin of extremely anisotropic superconducting gap in nematic phase of FeSe Superconductor. *Phys. Rev. X* **8**, 031033 (2018).

7. Chen, H. Y., MacQuarrie, E. R. & Fuchs, G. D. Orbital state manipulation of a diamond nitrogen-vacancy center using a mechanical resonator. *Phys. Rev. Lett.* **120**, 167401 (2018).

8. D'Ambrosio, V., Nagali, E., Marrucci, L. & Sciarrino, F. Orbital angular momentum for quantum information processing. *Proceedings of SPIE* **8440**, 84400F (2012).

9. Brange, F., Malkoc, O. & Samuelsson, P. Subdecoherence time generation and detection of orbital entanglement in quantum dots. *Phys. Rev. Lett.* **114**, 176803 (2015).

10. Yamamoto, M. et al. Electrical control of a solid-state flying qubit. *Nat. Nano.* **7**, 247-251 (2012).

11. Bassett, L. C. et al. Ultrafast optical control of orbital and spin dynamics in a solid-state defect. *Science* **345**, 1333-1337 (2014).

12. Tafti, F. F. et al. Resistivity plateau and extreme magnetoresistance in LaSb. *Nat. Phys.* **12**, 272 (2016).

13. Kumar, N. et al. Observation of pseudo-two-dimensional electron transport in the rock salt-

type topological semimetal LaBi. *Phys. Rev. B* **93**, 241106(R) (2016).

14. Nayak, J. et al. Multiple Dirac cones at the surface of the topological metal LaBi. *Nature Commun*. **8**, 13942 (2017).

15. Wu, Y. et al. Asymmetric mass acquisition in LaBi - a new topological semimetal candidate. *Phys. Rev. B* **94**, 081108(R) (2016).

16. Wu, Y. et al. Electronic structure of RSb (R = Y, Ce, Gd, Dy, Ho, Tm, Lu) studied by angle-resolved photoemission spectroscopy. *Phys. Rev. B* **96**, 035134 (2017).

17. Oinuma, H. et al. Three-dimensional band structure of LaSb and CeSb: Absence of band inversion. *Phys. Rev. B* **96**, 041120(R) (2017).

18. Guo, C. et al. Possible Weyl fermions in the magnetic Kondo system CeSb. *npj Quantum Mater.* **2**, 39 (2017).

19. Wang, Y. et al. Topological semimetal state and field-induced Fermi surface reconstruction in the antiferromagnetic monopnictide NdSb. *Phys. Rev. B* **97**, 115133 (2018).

20. Zeng, L. -K. et al. Compensated semimetal LaSb with unsaturated magnetoresistance. *Phys. Rev. Lett.* **117**, 127204 (2016).

21. He, J. F. et al. Distinct electronic structure for the extreme magnetoresistance in YSb. *Phys. Rev. Lett.* **117**, 267201 (2016).

22. Pavlosiuk, O., Swatek, P. & Wiśniewski, P. Giant magnetoresistance, three dimensional Fermi surface and origin of resistivity plateau in YSb semimetal. *Sci. Rep.* **6**, 38691 (2016).

23. Wakeham, N., Bauer, E. D., Neupane, M. & Ronning, F. Large magnetoresistance in the antiferromagnetic semimetal NdSb. *Phys. Rev. B* **93**, 205152 (2016).

24. Han, F. et al. Separation of electron and hole dynamics in the semimetal LaSb. *Phys. Rev. B* **96**, 125112 (2017).

25. Xu, J. et al. Origin of the extremely large magnetoresistance in the semimetal YSb. *Phys. Rev. B* **96**, 075159 (2017).

26. Zhou, Y. et al. Field-induced metamagnetic transition and nonsaturating magnetoresistance

in the antiferromagnetic semimetal NdSb. *Phys. Rev. B* **96**, 205122 (2017).

27. Liang, D. D. et al. Extreme magnetoresistance and Shubnikov-de Haas oscillations in ferromagnetic DySb. *APL Mater.* **6**, 086105 (2018).

28. Pavlosiuk, O., Swatek, P., Kaczorowski, D. & Wiśniewski, P. Magnetoresistance in LuBi and YBi semimetals due to nearly perfect carrier compensation. *Phys. Rev. B* **97**, 235132 (2018).

29. Hu, Y. J. et al. Extremely large magnetoresistance and the complete determination of the Fermi surface topology in the semimetal ScSb. *Phys. Rev. B* **98**, 035133 (2018).

30. Wang, Y. -Y. et al. Unusual magnetotransport in holmium monoantimonide. *Phys. Rev. B* **98**, 045137 (2018).

31. Yang, H.-Y. et al. Interplay of magnetism and transport in HoBi. *Phys. Rev. B* **98**, 045136 (2018).

32. Iwasa, K., Hannan, A., Kohgi, M. & Suzuki, T. Direct observation of the modulation of the 4$f$-electron orbital state by strong $p$-$f$ mixing in CeSb. *Phys. Rev. Lett.* **88**, 207201 (2002).

33. Wiener, T.A. & Canfield, P.C. Magnetic phase diagram of flux-grown single crystals of CeSb. *J. Alloys Compd.* **303–304**, 505–508 (2000).

34. Hulliger, F., Landolt, M., Ott, H. R. & Schmelczer, R. Low-temperature magnetic phase transitions of CeBi and CeSb. *J. Low Temp. Phys.* **20**, 269-284 (1975).

35. Takatsu, H. et al. Extremely large magnetoresistance in the nonmagnetic metal $PdCoO_2$. *Phys. Rev. Lett.* **111**, 056601 (2013).

36. Bogdanov, A. N., Zhuravlev, A. V. & Rößler, U. K. Spin-flop transition in uniaxial antiferromagnets: Magnetic phases, reorientation effects, and multidomain states. *Phys. Rev. B* **75**, 094425 (2007).

37. Welp, U. et al. Direct imaging if the first-order-spin-flop transition in the layered manganite $La_{1.4}Sr_{1.6}Mn_2O_7$. *Phys. Rev. Lett.* **96**, 4180-4183 (1999).

38. Xu, J. et al. Reentrant metallic behavior in the Weyl semimetal NbP. *Phys. Rev. B* **96**, 115152 (2017).

39. Park, B. G. et al. A spin-valve-like magnetoresistance of an antiferromagnet-based tunnel junction. *Nat. Mater.* **10**, 347-351 (2011).

40. Singh, A. et al. Colossal proximity effect in a superconducting triplet spin valve based on the half-metallic ferromagnet $CrO_2$. *Phys. Rev. X* **5**, 021019 (2015).

41. Kasuya, T., Sera, M., Okayama, Y. & Haga, Y. Normal and anomalous Hall effect in CeSb and CeBi. *J. Phys. Soc. Japan* **65**, 160-171 (1995).

42. Terashima, T. et al. Successive metamagnetic transitions and magnetoresistance in the low-carrier-density strongly correlated electron system CeP. *Phys. Rev. B* **58**, 309-313 (1998).

43. Kuroda, K. et al. Experimental determination of the topological phase diagram in cerium monopnictides. *Phys. Rev. Lett.* **120**, 086402 (2018).

44. Duan, C. -G. et al. Electronic, magnetic and transport properties of rare-earth monopnictides. *J. Phys.: Condens. Matter* **19**, 315220 (2007).

45. Lavrov, A. N. et al. Spin-flop transition and the anisotropic magnetoresistance of $Pr_{1.3-x}La_{0.7}Ce_xCuO_4$: Unexpectedly strong spin-charge coupling in the electron-doped cuprates. *Phys. Rev. Lett.* **92**, 227003 (2004).

46. Gnida, D. et al. Metamagnetism in $CePd_5Ge_3$. *J. Phys.: Condens. Matter* **25**, 126001 (2013).

47. Yokosuk, M. O. et al. Magnetoelectric coupling through the spin flop transition in $Ni_3TeO_6$. *Phys. Rev. Lett.* **117**, 147402 (2016).

48. Machado, F. L. A. et al. Spin-flop transition in the easy-plane antiferromagnet nickel oxide. *Phys. Rev. B* **95**, 104418 (2017).

49. Canfield, P. C. & Fisk, Z. Growth of single crystals from metallic fluxes. *Philos. Mag. Part B* **65**, 1117-1123 (1992).

**Figure captions**

**Fig. 1 | Magnetic-field dependent anisotropic magnetoresistance. a**, Schematic showing the definition of the angle $\theta$ for the magnetic field orientation. The magnetic field is rotated in the (100) plane while the current flows along the [100] direction, i.e. they are always perpendicular to each other. **b**, Polar plot of the angle dependent resistance of Sample A at $T = 3$ K in magnetic fields of $\mu_0 H = 9$ T to 2 T in intervals of 1 T. The data clearly show the minima in the ferromagnetic phase at $\mu_0 H \geq 5$ T at $\theta = 45°$, 135°, 225° and 315° degrees induced by orbital-flop, in addition to the minima at $\theta = 0°$, 90°, 180° and 270° degrees originating from the Fermi surfaces. **c**, Polar plot of the anisotropic magnetoresistances of the same sample at the same temperature in antiferromagnetic phase from $\mu_0 H = 4$ T to 1 T (and partial data at 4.5 T) in intervals of 0.5 T. The data show more complicated features resulting from the competition of orbital-flop, anisotropic Fermi surface and magnetic phase transitions. **d**, Schematic showing the orientations of the Ce $\Gamma_8^{(1)}$ orbitals (light blue crosses) that flops at $\theta = 45°$, 135°, 225° and 315° degrees. Black arrows represent the magnetic moment directions.

**Fig. 2 | Anisotropy of the ferromagnetic transition. a**, $R(H)$ curves of Sample A at $T = 3$ K and $\theta = 90°$, 135° and 180°. $H_{FM}$ is the magnetic field above which the system is in the ferromagnetic state. $H_{FM*}$ indicates a possible phase transition. $H_{CR}$ is the magnetic field above which the magnetoresistance at **H** ∥ [011] is smaller than that at **H** ∥ [010] and **H** ∥ [00$\bar{1}$]. **b**, Angle dependence of $H_{FM}$ (solid symbols) and $H_{FM*}$ (open symbols). The dashed lines represent $1/cos\varphi$ with $\varphi = \theta - n\pi/2$ with $n = 1$, 2, and 3 for $90° \leq \theta \leq 135°$ and $135° \leq \theta \leq 225°$, and $225° \leq \theta \leq 270°$, respectively (see text for more discussion). **c**, Phase diagrams $H_{FM}$ versus $T$ for the ferromagnetic states at **H** ∥ [010] and **H** ∥ [011] and for $H_{CR}$, above which orbital-flops appear as

clear magnetoresistance minima in the $R(\theta)$ curves. Results indicate that orbital-flop effect is most pronounced in the ferromagnetic phase. **d**, Magnetoresistances at $\mu_0 H = 7$ T and temperatures from 3 K to 25 K. Orbital-flop induced minima can be seen at $\theta = 45°$, $135°$, $225°$ and $315°$, i.e., **H** ∥ <011> at $T < 10$ K.

**Fig. 3 | Anisotropic magnetization induced by orbital flops**. **a**, Angle dependence of the measured magnetization of Sample C at $T = 5$ K and at various magnetic fields. **b**, Angle dependence of the measured magnetization at $\mu_0 H = 7$ T and at various temperatures. **c**, Temperature dependence of the measured magnetization at $\mu_0 H = 7$ T and at two magnetic field orientations. $\gamma$ is the ratio of the magnetizations at [010] and [011]. The cyan curves in (b) represent theoretical results from the phenomenological model. **d**, Angle dependence of the magnetizations along the [010] (**b**-axis) and [001] (**c**-axis) derived from a phenomenological model (see text) for the ferromagnetic phase, revealing an orbital-flop at $\theta = 45°$. $\mu_B$ is the Bohr magneton.

**Fig. 4 | Probing orbital flops using a spin-valve like structure**. **a**, Schematics of the structure of Sample B (see Supplementary Figure 9 for an optic image) and the definition of $\theta$ for the magnetic field direction. This sample has only two contacts of ferromagnetic permalloy (Py) fabricated via sputtering-deposition. The interface between the Py contact layer and the CeSb crystal forms a spin-valve like structure. The magnetic field is rotated in the same (100) plane as that for Sample A. **b**, Equivalent circuit for Sample B. $R_{SV}$ and $R_{CeSb}$ represent the resistances at the Py/CeSb interface and in the CeSb crystal, respectively. **c**, Orientations of the magnetic moments in the Py layer and in the CeSb crystal at various orientations of an applied high magnetic field (e.g. $\geq 5$ T). A schematic of the expected angle dependence of the interface resistance $R_{SV}$. **d**, Angle dependence of the measured total resistance $R = 2R_{SV} + R_{CeSb}$ of Sample B at $T = 3$ K and

various magnetic fields/orientation. The spin-valve like behavior described in (c) is reflected by the significant suppression of the magnetoresistances at $45° < \theta < 135°$ and $225° < \theta < 135°$ compared to those at $0° \leq \theta < 45°$, $135° < \theta < 225°$ and $315° < \theta \leq 360°$.

# Figure 1

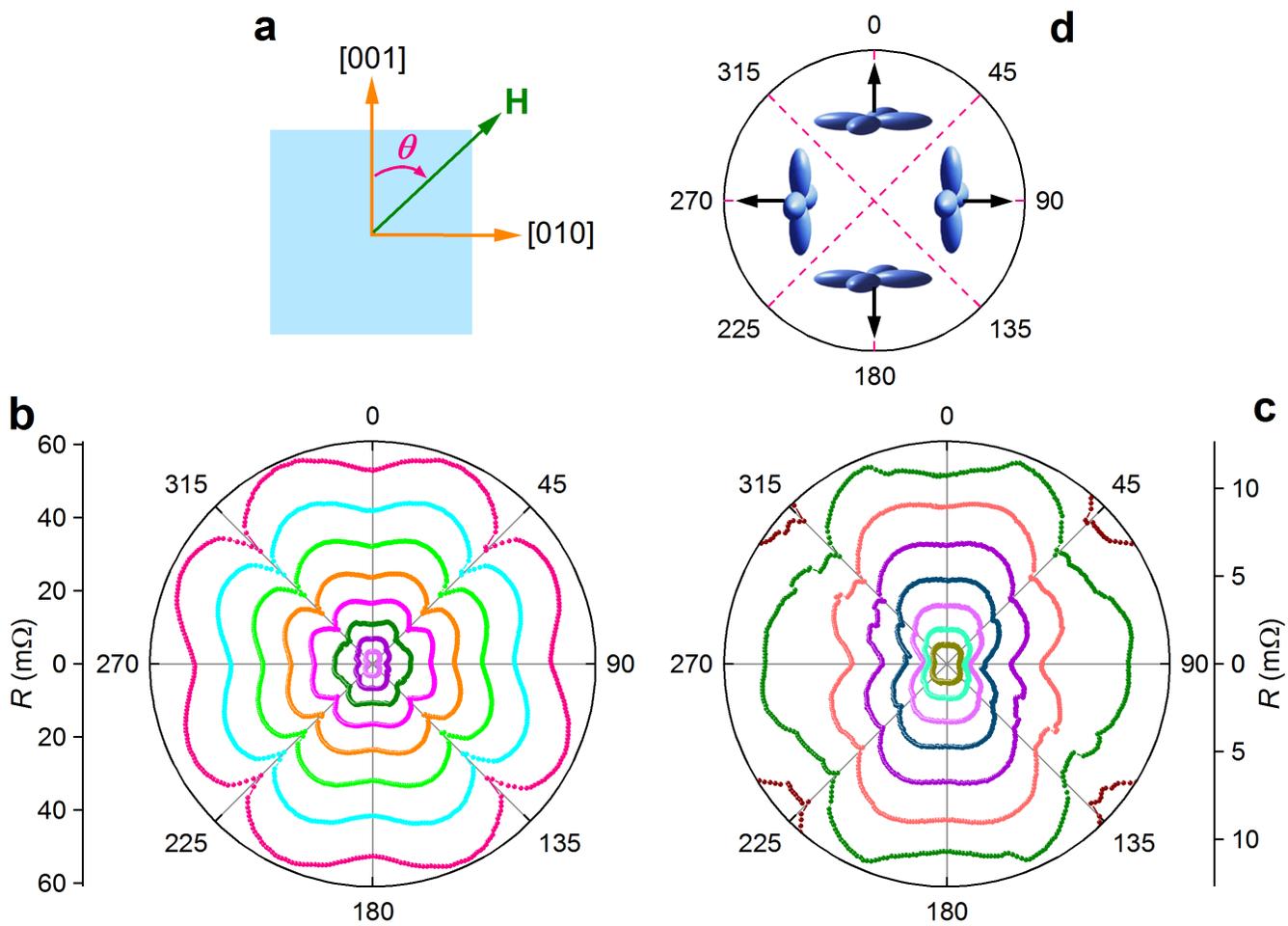

**Figure 2**

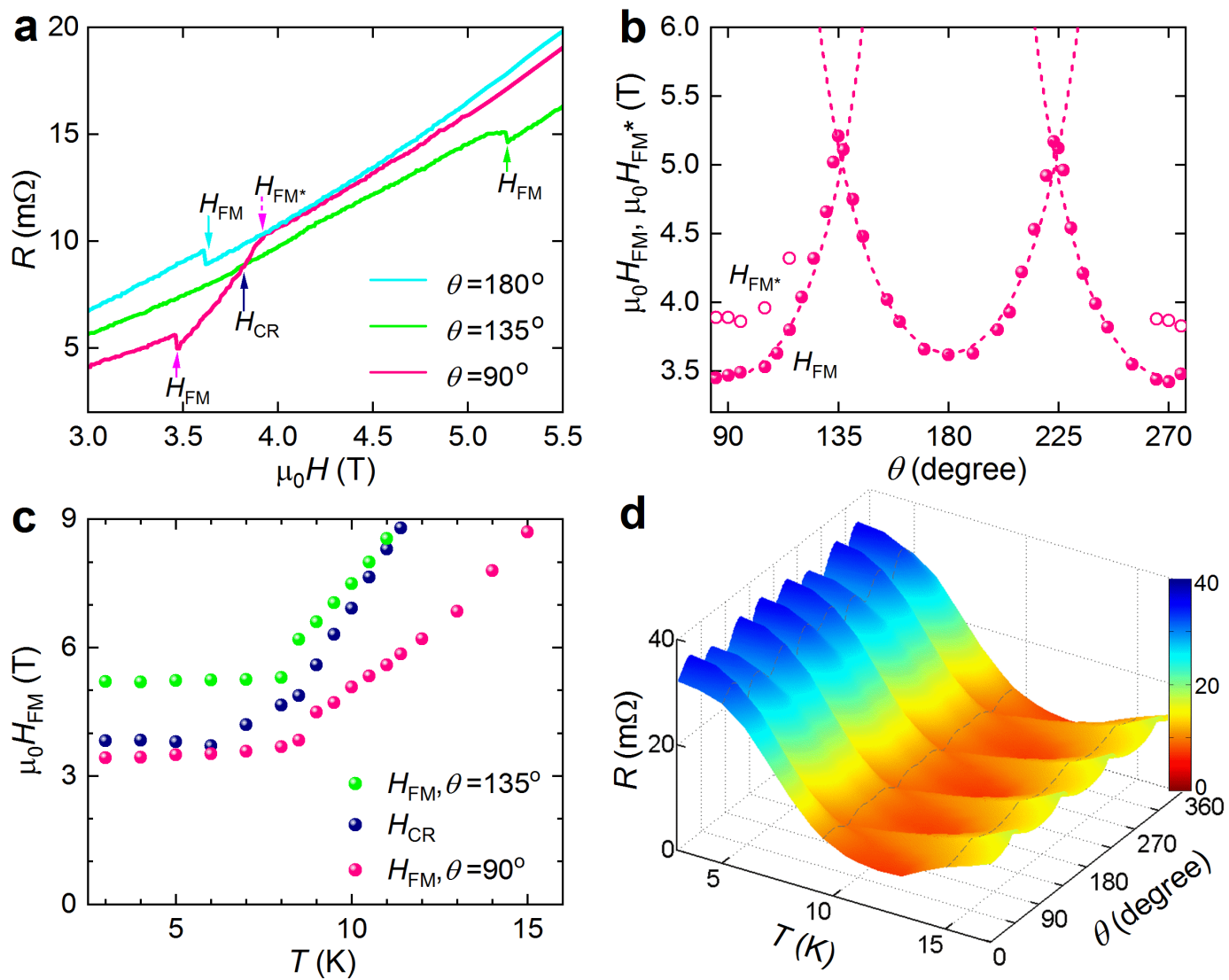

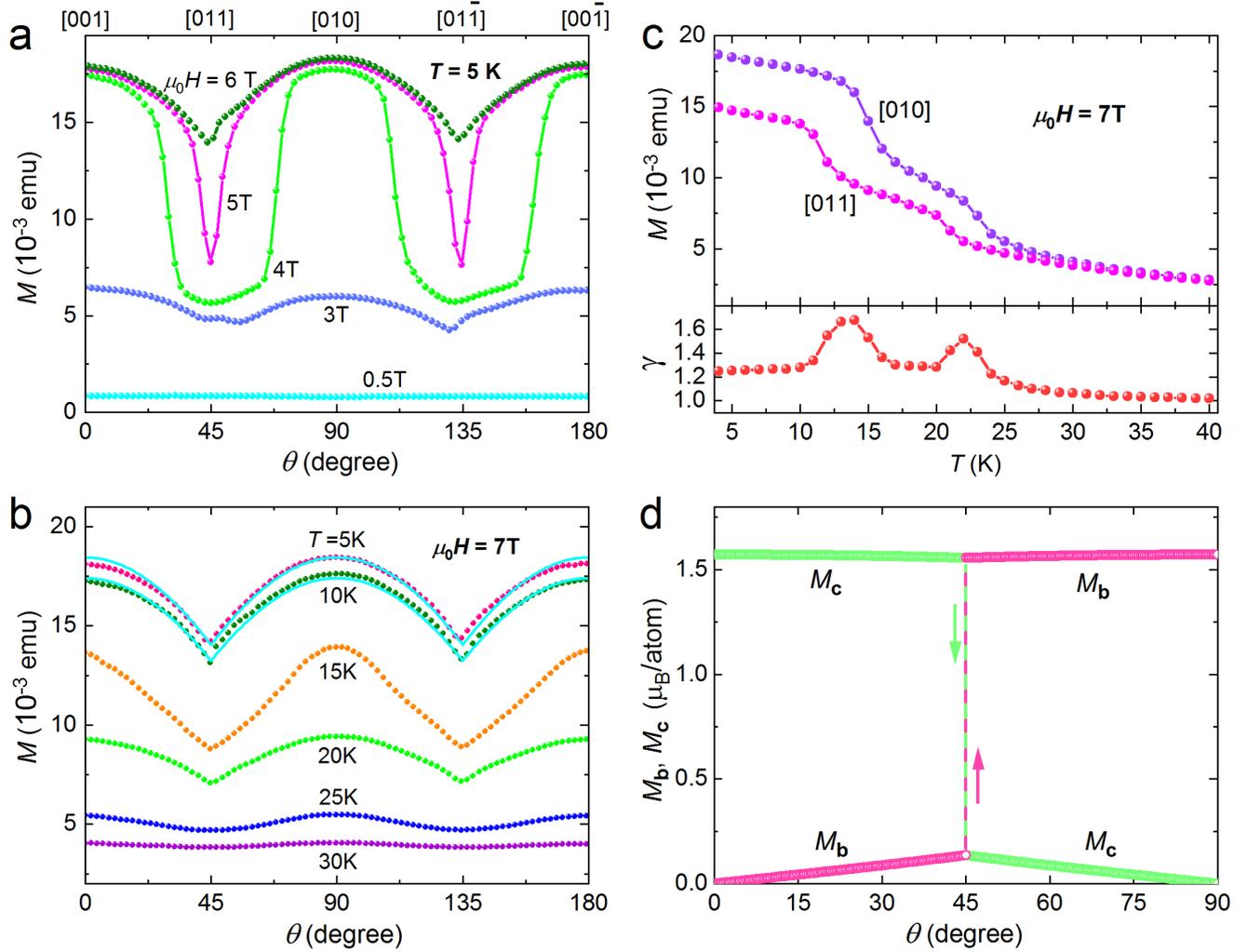

**Figure 4**

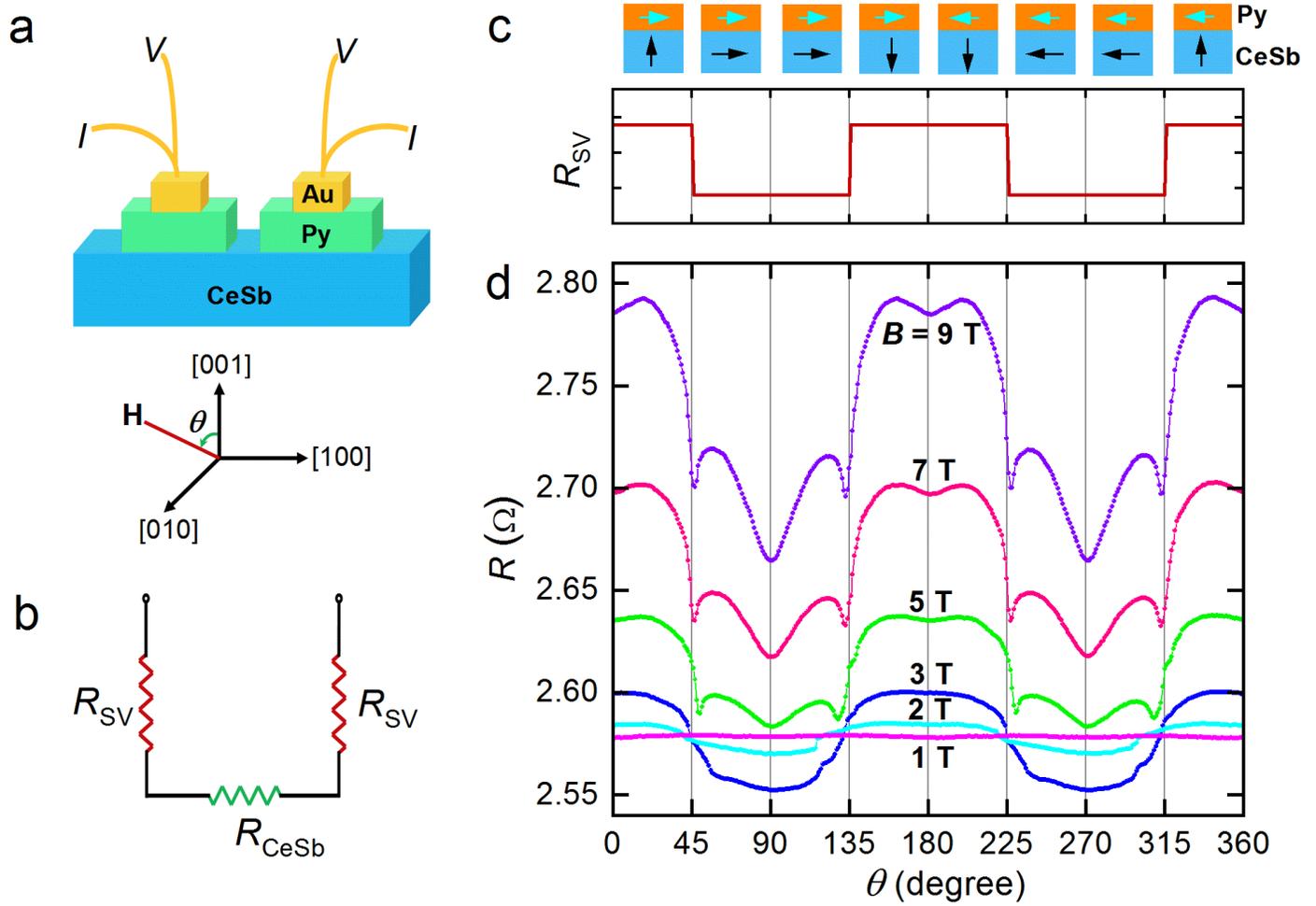